\documentclass{elsart}
\usepackage{graphicx}

\begin{document}

\begin{frontmatter}
\title{Epidemics, disorder, and percolation}  
\author{\corauthref{cor1}L. M. Sander,}
\corauth[cor1]{Corresponding author}
\author{C. P. Warren}
\address{Michigan Center for Theoretical Physics, University of Michigan, Ann Arbor, 
Michigan - 48109, U.S.A.}
\author{I. M. Sokolov}
\address{Institut f\"ur Physik, Humboldt-Universit\"at zu Berlin, Germany
Invalidenstr. 110, 10115 Berlin}

\begin{abstract} 
Spatial models for spread of an epidemic may be mapped onto bond percolation. 
We point out that with disorder in the strength of contacts between 
individuals patchiness in the spread of the
epidemic is very likely, and the criterion for epidemic outbreak depends
strongly on the disorder because the critical region of the
corresponding percolation model is broadened. 
In some networks the percolation threshold is zero if 
another kind of disorder is present, namely divergent
fluctuations in the number of contacts. We give an example,
a network with a well defined geography,  where this is
not necessarily so, and  discuss whether real infection networks are likely
to have this property.
\end{abstract}

\end{frontmatter}

\section{Introduction}
Models for  the spread of epidemics are an interesting application of
non-equilibrium statistical physics. In this paper we discuss two aspects of this
sort of study in cases when disorder is present. First we generalize the
well-known \cite {Grassberger} application of percolation theory to epidemics
for which variation in the strength of coupling. Second, we consider the contact
networks which carry epidemics.  If there is large variation in the number of
contacts between members of the population, then the network structure can lead
to a  percolation threshold of zero, for example if the distribution of contacts
is a suitable power law.  We show an example of a network which seems to have
$p_c \ne 0$ for the same power-law distribution, and which might have some
relevance to real infections. 

\section{SIR epidemics in an inhomogeneous population}
The SIR (Susceptible, Infected, Recovered) model is a classic way to think about
the spead of epidemic infections \cite{Murray}. It is based on the idea that in
many epidemics an individual catches a disease, infects others at rate $x$ for a
certain period, $\tau$, and then recovers or dies, and is thus removed from the
infection network. 
A mean-field version of this
model is given by the SIR equations\cite{Murray} for a perfectly mixed
population:
\begin{equation}
dS/dt = -xSI/N, \qquad
dI/dt = xSI/N - I/\tau
\label{SIR}
\end{equation}
Here N is the size of the population, and we normalize $x$ in this
way so that we can think about the infection rate per contact. Clearly,
we have an outbreak if $R_o = xS(0)\tau/N > 1$. 

We can think of Eq. (\ref{SIR}) as describing dynamics on a fully
conjugated graph -- every node sees every other equally. In this paper  we 
will consider a model  on a lattice, where each individual only sees its neighbors.
Each site of the lattice is initially occupied by a susceptible ($S$) individual. One lattice site
is initially infected, $I$. Any $S$ can be infected by $I$'s which are
nearest neighbors. The probability of infection per unit time along the 
$i^{th}$ bond, $x_{i}$, is chosen from a distribution $f(x)$. 
For  a case near the threshold for the epidemic (see below) the pattern of infection is shown in 
Figure \ref{snap}. The similarity of this pattern to a percolation cluster is very clear.  

\begin{figure}
\parbox{2.5in} 
{\includegraphics[width=2.5in]{fig1.EPS} 
\caption{ Pattern of recovered after the epidemic
has died out on a 256x256 lattice. 
Recovered sites are gray, and uninfected black. } 
\label{snap}}
\hspace{.25in}
\parbox{2.5in}
{\includegraphics[width=2.5in, height=2.3in]{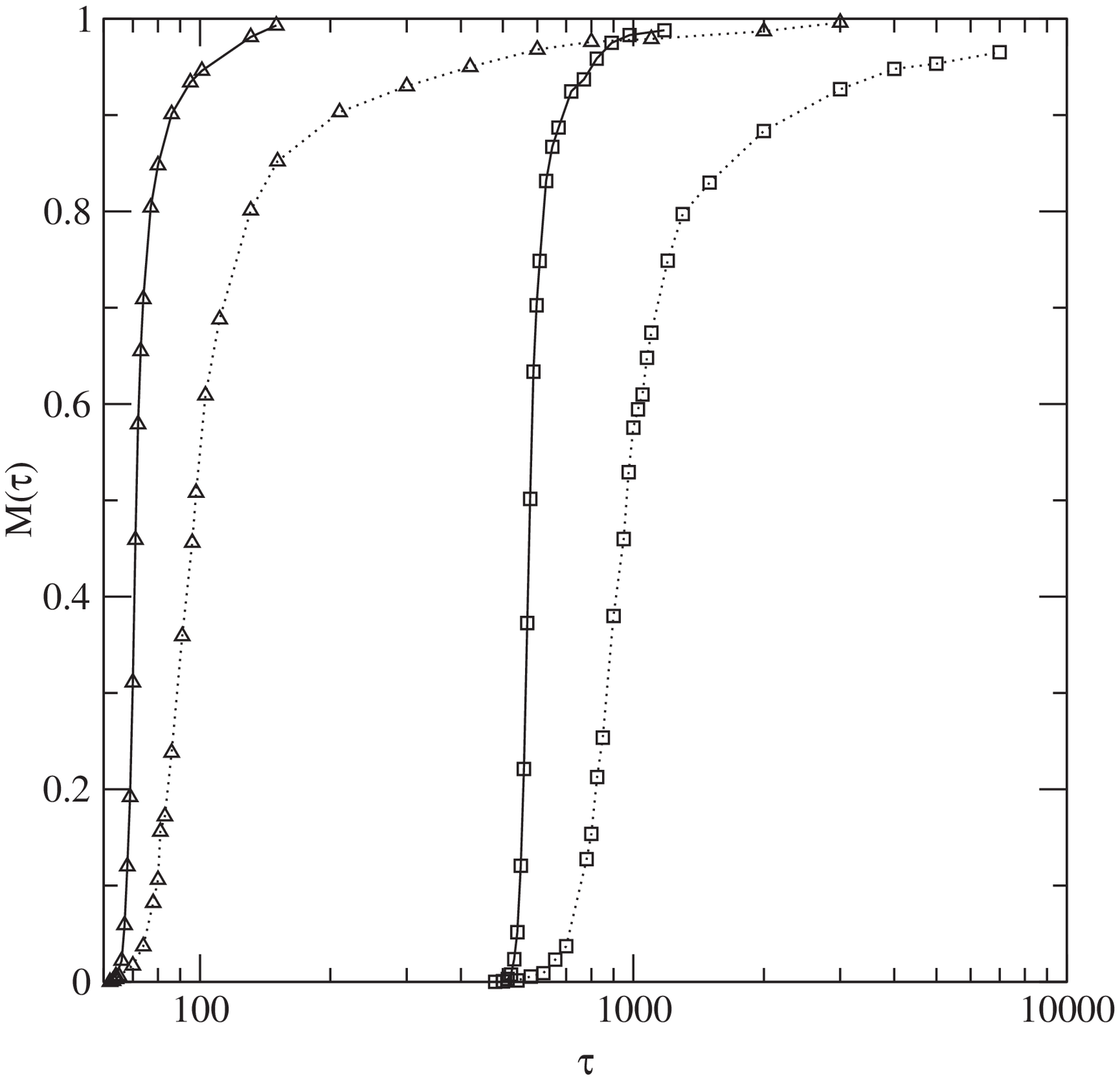}
\caption{$M(\tau)$ for a triangular lattice, 
$\bigtriangleup$, and a square lattice, $\Box$. Solid lines:$f(x)$ from
Eq.~(\ref{weak}) $x_{max}=0.003$. Dotted lines: Eq.~(\ref{strong}) with $x_{min}=e^{-15}$.
 }
\label{M(tau)}}
\end{figure}  
  
The existence of a large epidemic depends on the parameters just as in the mean-field case. 
For fixed $x_i$, if the recovery time, $\tau$, is too
small the epidemic will die out, and if $\tau$ is large enough it will persist.
A critical value, $\tau_c$  marks the threshold.  However, even for
large $\tau$ there is a finite probability that the epidemic dies. We define a
spanning probability, $M(\tau)$, the probability that an infection started from a single site
in the middle of a large lattice reaches the edge. $M$ grows rapidly for $\tau > \tau_c$.
In Figure \ref{M(tau)} we show $M$ for a 200 x 200 lattice as a
function of $\tau$  for two different choices of
$f(x)$, one rectangular, and one strongly asymmetric.
 
This model is not obviously related to percolation theory. 
For  a very narrow
distribution, $f(x) = \delta (x-x_{o})$ Grassberger 
\cite{Grassberger}
showed that we do have percolation as follows: the probability to infect a neighbor is 
$p(\tau )=1-(1-x_{o})^{\tau }$. Thus $p$ is the
fraction of bonds completed in a given epidemic, and the infection
process is mapped onto bond percolation with $p(\tau )$ playing the
role of the percolation probability. For a square lattice in two
dimensions, $p_{c}=1/2$, so we expect that when $p$ is greater than
$1/2$ we will have an epidemic. This gives a critical recovery time:
$\tau _{c}=-\ln 2/\ln (1-x_{o})$. For any
other network, 
$p_{c}= 1-(1-x_o)^{\tau_c}$
determines the threshold for epidemic spread.
For arbitrary $f$'s, we can write: 
\begin{equation}
p(\tau )=1-\int (1-x)^{\tau }f(x)dx = p_c. 
 \label{prob}
\end{equation}
Thus, for any $f$ we have mapped our infection problem to percolation. The
solution of Eq. (\ref{prob})
gives  $\tau_c$. The mapping of a dynamic problem onto static
percolation is not trivial: different runs for infection spreading in a
system with a given realization of $x_{i}$ give different, but not independent
realizations of a percolation problem, i.e. of a set of a completed bonds. In previous work
\cite{mathbi} we have verified Eq. (\ref{prob}) by doing a data collapse for various different
choices of $f(x)$. After data collapse, the  function $M$, expressed as a function 
of $p$ using Eq.  (\ref{prob}) is identical to the mass of the infinite cluster 
in ordinary percolation theory.
All of the results of this work may be applied easily for any network. 
 
In order to understand the effect of the distribution of bond strengths
we studied $f$'s which are either weakly or
strongly asymmetric.
An example of a symmetric $f$ is:
\begin{equation}
f_w(x) = 1/x_{max}, \quad 0\le x \le x_{max}. 
\label{weak}
\end{equation}
In cases like this  where $f(x)$ is concentrated near its
mean value,
it is easy to show 
the spread of $x_i$ does
influence the infection propagation very much.

However, the dependence $p(\tau)$, Eq.(\ref{prob}), is strongly nonlinear and
is dominated by the behavior of $f(x)$ in vicinity of $x=0$: the 
spread of the infection is controlled by
weakest bonds. Thus strongly asymmetric distributions
concentrated near zero should be very different from the case 
discussed above. Consider, for example, power-laws
$f(x)=(\alpha -1)x^{-\alpha }$
with $0<\alpha <1$.  In this case $\int (1-x)^{\tau }f(x)dx=(\alpha
-1)B(\tau +1,1-\alpha )$, where $B(x,y)$ is the
beta-function. Strongly asymmetric distributions correspond to values
of $\alpha $ in vicinity of $1$. Using the fact that for such distributions
$\tau \gg 1$, we can use the Stirling formula to get in 
leading order:
$\tau _{c}\simeq \left[ \Gamma (2-\alpha )(1-p_{c})\right]
^{-1/(1-\alpha )}$
which diverges for $\alpha \rightarrow 1$.  

An even more extreme case is that of
functions $f(x)$  with $\alpha \ge 1$.
These  are not acceptable probability densities
 since they are not normalizable on [0,1]. However, one can truncate
the $f$ in the vicinity of zero to allow normalization. For example:
\begin{equation}
f_{s}(x)=C/x\quad x_{min}\le x\le 1;\quad C=1/|\ln (x_{min})|.
\label{strong}
\end{equation}
The two distributions considered in Figure \ref{M(tau)} are given by 
Eq. (\ref{strong}) with $x_{min}= e^{-15}$, an extremely asymmetric case, 
and Eq.(\ref{weak}), the symmetric case. 

In Figure \ref{M(tau)} we see that the effect of strong asymmetry is to
spread out the critical region, i.e. the region where $M$ is not close to either
1 or 0. This
is the parameter region where the epidemic pattern is patchy, as in Figure
\ref{snap}. For a strongly asymmetric $f$
the patchy pattern  corresponds to a large range of the
parameter $\tau$. We recall from
percolation theory that the regime where $M$ is neither very small nor very
close to 1 is where the fractal nature of
the percolation cluster at $p_c$ persists over large length scales. This could
have implications for practical epidemiology: if the actual distribution of
$x_i$ is very asymmetric, we would be very likely to observe patchiness. For
such distributions mean-field theory is very inaccurate \cite{mathbi} and
sampling of populations must be done with care because of the spatial
correlations.

\section{The network of contacts: power-law distributions}
In the previous section we considered networks for which
the number of contacts between
individuals does not vary much, though we supposed that the strength
of such contacts varied a good deal. However, there is data to suggest
\cite{Amaral} that some infection networks have a wide distribution of the
number of contacts, so that the bond number, $k$, is drawn from a
distribution of the form $P(k) \propto 1/k^{\alpha}$ with $ 2 <\alpha <3$. In
this case the mean value of the number of contacts, $<k>$, exists, but the
fluctuation, $<k^2>$ diverges.

We have seen that it is sufficient to consider percolation 
on such networks to understand SIR epidemics.
This problem has been discussed a good deal
\cite{HavlinAlpha3,cnsw,Vazquez}. There is general agreement that $p_c$ is zero
for many cases. The graphs considered by these authors are essentially
random graphs with a given degree distribution. Thus the model has a kind
of random mixing.

The random graphs dominate the the ensemble of all graphs with a given
$P(k)$. The study of this ensemble was introduced
by Molloy and Reed \cite{Molloy-Reed} and a criterion for the formation of a
giant component was derived. Callaway, et. al 
\cite{cnsw} and Cohen, et al. \cite{HavlinAlpha3} express a related result for 
percolation: 
$p_c ( <k^2>/<k>^2 -2) =1$
Thus if $<k^2>$ diverges we expect $p_c=0$. 
We can see this by starting at a random node on
the graph so that there are $k_0$ bonds of which $pk_0$ are active. 
If neighbors are chosen at random, the probability for the number of bonds on the next 
neighbor is $p^2 k_1 P(k_1)/<k>$. The factor of $k_1$ comes from the fact that highly bonded
sites are more likely to have a bond with the first site. Now the number of infected increases 
by a factor $<k_1>/<k_0> = p<k^2>/<k>^2$ at each such step. 
If $<k^2>$ is finite, then this can be made less than unity for  $p$ small.
and the infection will die out. If $<k^2>$ is infinite, then $p$ must be 0.

\begin{figure}
\centerline{\includegraphics*[width=2.in,height=2.in]{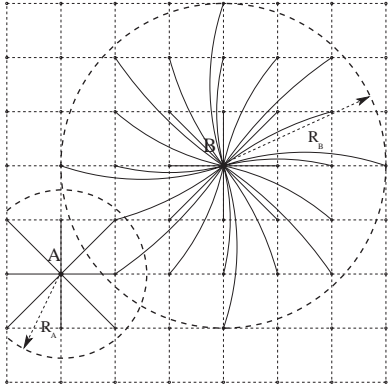}}
\caption{The geographic model:
One disk with radius $R_A$ is centered around node $A$,
and another with radius $R_B$ is centered about $B$.  All nodes within each disk are attached to
the central node.}
\label{Disk}
\end{figure}

Now consider a model which has the same $P(k)$ but has \emph{geography} \cite{SSWcond,Havl}.
We assume that individuals infect their neighbors, but that the number of infections is
drawn from $P$. 
We construct this by taking each point on a $d$-dimensional lattice with probability $p$
and connecting them
all other points within a $d$-sphere of volume $k$, centered on the point. In 2 dimensions 
we surround each point with a disk of radius of roughly $R=[k/\pi]^{1/2}$. The probability 
distribution of $R$ is given by $P(R) = P(k) dk/dR \propto R^{-(d\alpha -d+1)}$.
Our model is a variant of an
old two-dimensional circle model of continuum percolation 
\cite{Contperc}
with variable radius of circles.
See Figure \ref{Disk}.
This is a small part of the Molloy-Reed ensemble, but we suggest that it may be relevant for real-world epidemics: sometimes infection is primarily local.

Note that a site can have the bonds it generates itself (`proper' bonds) and
those from other disks. 
We show that the total number of bonds per node has
the same power-law behavior as the
proper bonds if $\alpha>2$ by calculating the mean number of bonds connecting a
node $i$ to nodes $j$ whose center is a distance $r >R_i$ away. 
The probability to find $R_j$ larger than $r$ is 
$\int_{r}^{\infty }P(r^{\prime })dr^{\prime }$.  Thus, 
the mean number of nodes connected to $i$ from outside is 
$
k_{out}\propto \int_{R_{i}}^{\infty}dr r^{d-1} \int_{r}^{\infty }P(R_{j})dR_{j}.
$
Thus
$k_{out} \simeq R_{i}^{-d(\alpha -2)}$
which tends to zero for large $R$, as long as  $\alpha >2$. This means that
the probability distribution of the number of the `proper'  bonds and the
actual number of the bonds of a node is the same for $k$ large, and that 
bonds from outside are rare.
  
For $d=1$ it is easy to see that  the system never percolates,
because for any $p<1$ there will eventually be a gap of length $\Delta$
in the chain of connected disks.  Consider  disks  with radii between $k$ and  $k+dk$;
the concentration of their centers is $pP(k)dk$. The probability
that the gap is not covered by one of these disks is equal to
the probability that none of their centers are found in the
interval of length $2k+\Delta$ (in units of the lattice constant) centered at the gap,
$1- pP(k)(2k+\Delta)dk \simeq \exp[-pP(k)(2k+\Delta)dk]$
 The overall probability that no disk overlaps the gap is the product over $k$:
\begin{equation} 
\exp \left(- \int _{0} ^ {\infty}p  P(k)(2k+\Delta)dk\right)=
\exp \left(- 2p \left< k\right>-p\Delta \right),
\end{equation}
which is finite as long as $\left<k \right>$ exists.

\begin{figure}
\centerline{\includegraphics*[width=4in, height=3in]{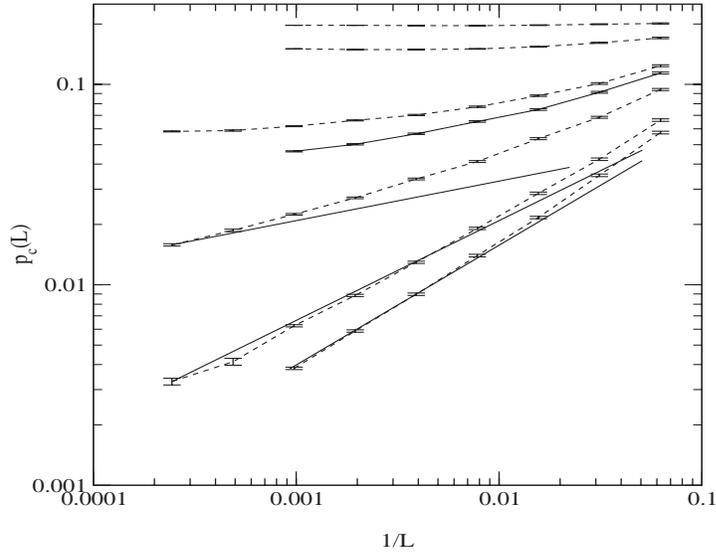}}
\caption{
$p_c$ on an $L$ x $L$ lattice as a function of  $1/L$. 
From top to bottom, the data are for $\alpha=2.5, 2.3, 2.05$ (dotted lines),
 $2$, (solid line), and
$1.95, 1.9,$  and $1.7$ (dashed lines).  The solid
straight lines have slope $2(\alpha-2)$.}
\label{f2}
\end{figure} 
In 2  dimensions we did simulations
to find $p_c$.  Disks were randomly added to an $L$ x $L$ lattice at different
sites until a cluster spanned the lattice.  Figure \ref{f2} shows the average percolation 
threshold $p_c(L)$ as a function of $1/L$.  It appears from the plot that 
there is a finite percolation threshold for not just $\alpha>3$ but $2<\alpha<3$ as well, 
the region of interest for real world epidemics.   For $\alpha<2$, the results are consistent with $p_c=0$ and seem to scale
according as $L^{2(\alpha-2)}$ for sufficiently large $L$.  We have shown \cite{SSWcond}
that this scaling is related to the presence of giant disks that span the whole sample. These
occur with finite probability for $\alpha < 2$.

We can try to understand the results in Figure \ref{f2} by  calculating $<k_1>$ as above. 
Most of the bonds emanating from the center of  disk point within it. These 
have degrees drawn from the distribution $P(k_1)$, not  $k_1P(k_1)$. Thus, for the
typical behavior we $<k_1> = <k>$ for $\alpha >2$. However, there are rare events which
correspond to the overlap of large disks from outside with the center of the starting disk. These
give a large contribution. In fact, in 2 dimensions for a disk of size $R_0$ we add up the contributions of outside disks of radius $R_1$:
\begin{equation}
<k_1> \propto  \int_{R_0}^{\infty} r dr  \int_{r}^{\infty} R_1^{2}P(R_1)dR_1 = 
 \int_{R_0}^{\infty} dr r^{-(2\alpha -5)} 
\end{equation}
 This diverges for $\alpha < 3$.  Note that this is an overestimate since have not attempted to exclude already infected sites.

We have a confusing situation: the typical behavior is that the
cluster of disks is finite in extent, but the average mass added diverges. There is a rigorous
theorem in the study of continuum percolation which treats exactly this situation \cite{Contperc}.
It states that if $<k^{2-1/d}>$  is finite, but $<k^{2}> $ diverges ( in 2 dimensions $2.5 < \alpha <3$),
then there is  a $p_c$ such that for $p<p_c$ \emph{with probability unity all clusters are finite, but the 
expected number of points of the largest cluster is infinite.} (The probability to span a large
lattice is not addressed by this theorem.)
We interpret this odd situation, at least for $\alpha > 2.5$ as a form of intermittancy. Typical epidemics are finite for small $p$, but  huge, rare events occur. It is interesting to 
ask the physical question of what we would expect to observe in this case. We claim (perhaps
too hopefully) that the situation is best represented by the computer simulations. Figure \ref{f2} 
shows that at $\alpha = 2.5$ there is quite solid evidence for a non-zero $p_c$, and this
seems to extend down to $\alpha =2$. (Note that the theorem does not exclude this possibility). 

In summary we have a counterexample to the prevailing notion that $p_c$ must be zero if
$<k^2>$ is infinite, certainly in one dimension. In 2 dimensions the situation is complex, and
will take more  work to fully sort out.

\begin{ack}
We have had many useful conversations with Mark Newman and Carl Simon.
\end{ack}

\end{document}